%% file: main.tex
\documentclass[conference,10pt]{IEEEtran}

\makeatletter\def\input@path{{current}}\makeatother

\input{auxiliary/packages}

\input{auxiliary/preamble}
\input{auxiliary/definitions}
\input{auxiliary/graphics}

\input{auxiliary/tables}
\IEEEoverridecommandlockouts
\begin{document}

\graphicspath{ {./figures/} }

\newcommand\combined{\hspace{-4.5mm} The views and
  conclusions contained in this document are those of the authors and
  should not be interpreted as representing the official policies,
  either expressed or implied, of the Army Research Laboratory or the
  U.S. Government. The U.S. Government is authorized to reproduce and
  distribute reprints for Government purposes notwithstanding any
  copyright notation herein.  This work was partially funded by Cyber
  Technologies, Deputy CTO for Critical Technologies/Applied
  Technology, Office of the Under Secretary of Defense Research and
  Engineering.  Dr.~Vandekerckhove's research was sponsored by the
  Army Research Laboratory Cooperative Agreement Number
  W911NF-21-2-0284 and National Science Foundation
  grants \#1850849 and \#2051186.  The authors would like to thank the anonymous reviewers for carefully
reading the paper and for providing helpful suggestions.}

\title{Piecewise Linear and Stochastic Models for the Analysis of Cyber Resilience\thanks{\combined}}

\author{\IEEEauthorblockN{Michael J. Weisman, Alexander Kott}
\IEEEauthorblockA{
  \textit{U.S. Army Combat Capabilities Development Command}\\
  \textit{Army Research Laboratory} \\
  Adelphi, MD \\
  \{michael.j.weisman2, alexander.kott1\}.civ@army.mil
}
\and
\IEEEauthorblockN{Joachim Vandekerckhove}
\IEEEauthorblockA{\textit{University of California, Irvine} \\
  \textit{Department of Cognitive Sciences}\\
  Irvine, CA \\
  joachim@uci.edu}
}

\maketitle

\input{includes/00-abstract.tex}
\input{includes/10-introduction-new}
\input{includes/20-prior-work-revisited}
\input{includes/40-continuous-model-new}
\input{includes/figure_3}

\input{includes/50-stochastic-differential-equation-model-new}

\input{includes/52a-piecewise-linear-sde}

\input{includes/60-relationship-between-ode-and-sde-new}
\input{includes/70a-generating-stochastic-realizations}
\input{includes/figure_4}

\input{includes/75-parameter-estimation.tex}

\input{includes/76-next-steps.tex}

\input{includes/77-linear-filtering.tex}
\input{includes/80-discussion-and-conclusion-new}

\bibliographystyle{IEEEtran}
\bibliography{references.bib}
\end{document}

%% file: auxiliary/packages.tex
\usepackage{amsmath,amssymb,amsfonts,amsthm}
\usepackage{algorithmic}
\usepackage{graphicx}
\usepackage{textcomp}
\usepackage{xcolor}
\usepackage[utf8]{inputenc}
\usepackage[margin=1in]{geometry}
\usepackage[T1]{fontenc}

\usepackage[english]{babel}
\usepackage[square, numbers]{natbib}                                                   
\usepackage{xfrac}

%% file: auxiliary/preamble.tex
\def\BibTeX{{\rm B\kern-.05em{\sc i\kern-.025em b}\kern-.08em
    T\kern-.1667em\lower.7ex\hbox{E}\kern-.125emX}}

%% file: auxiliary/definitions.tex
\newcommand{\blue}[1]{#1} 

\newcommand{\expminus}{e^{-\int_0^t \Qware(p) \, dp }}
\newcommand{\expplus}{e^{\int_0^\tau \Qware(p) \, dp }}
\newcommand{\bintegral}{\Fnominal \int_0^t \expplus \bonware(\tau) \, d \tau}  

\newcommand{\bonwareSuper}{b}
\newcommand{\malwareSuper}{m}

\newcommand{\functionality}{{{F}}}

\newcommand{\activity}{{{A}}}
\newcommand{\effectiveness}{{{E}}}

\newcommand{\malwareActivity}{{\activity}^{\malwareSuper}(t)}
\newcommand{\bonwareActivity}{{\activity}^{\bonwareSuper}(t)}
\newcommand{\malwareEffectiveness}{{\effectiveness}^{\malwareSuper}(t)}
\newcommand{\bonwareEffectiveness}{{\effectiveness}^{\bonwareSuper}(t)}

\newcommand{\malwareRate}{\theta^{\malwareSuper}}
\newcommand{\bonwareRate}{\theta^{\bonwareSuper}}
\newcommand{\malwareSize}{{{\gamma}^{\malwareSuper}}}
\newcommand{\bonwareSize}{{{\gamma}^{\bonwareSuper}}}

\newcommand\bonware{\mathcal{B}}
\newcommand\malware{\mathcal{M}}
\newcommand\Qware{\mathcal{Q}}

\newcommand\Fnominal{\blue{F_\text{N}}{}}

\DeclareMathOperator{\erf}{erf}
\DeclareMathOperator{\unif}{Unif}
\DeclareMathOperator{\bern}{Bern}

\newcommand{\dunif}[2]{\unif\!\left(#1,#2\right)}
\newcommand{\dbern}[1]{\bern\!\left(#1\right)}

\newcommand\impact{\blue{impact}} 

%% file: auxiliary/graphics.tex
\newcommand\fff{}

\newcommand\aaa[1]{\color{black}\fff{\includegraphics[width=0.98\columnwidth]{#1}}\vspace{-1mm}}
\newcommand\eee[1]{\color{black}\fff{\includegraphics[width=0.995\columnwidth]{#1}}\vspace{-1mm}}

%% file: includes/00-abstract.tex
\begin{abstract}
\label{abstract}
We model a vehicle equipped with an autonomous cyber-defense system 
in addition to its
inherent physical resilience features.  When attacked, this ensemble of cyber-physical features (i.e., ``bonware'') strives to resist and recover from the performance degradation caused by the malware's attack.  We model the underlying differential equations governing such attacks for piecewise linear characterizations of malware and bonware, develop a discrete time stochastic model, and show that averages of instantiations of the stochastic model approximate solutions to the continuous differential equation.  We develop a theory and methodology for approximating the parameters associated with these equations.
\end{abstract}

%% file: includes/10-introduction-new.tex
\section{Introduction}
In this paper, we report progress we have made on a project called \textit{Quantitative Measurement of Cyber Resilience} (QMoCR) whose goals include building a mathematical model to characterize cyber resilence, and finding objective quantitative measures of cyber resilience.  In \cite{kott2022}, we began to develop the tools to model cyber resilience mathematically.  In a companion paper \cite{ellis2022experimental}, we described an experimental testbed we have developed, and gave examples of the data we can produce.  In the current paper, we continue to develop the mathematical model.  In particular, we expand on the piecewise linear model we described in \cite{kott2022} and exhibit its stochastic counterpart.

\input{includes/cyber_resistence_summary}

%% file: includes/cyber_resistence_summary.tex
The QMoCR program has focused on two research areas: (1) mathematical modeling of cyber resilience and (2) framing an infrastructure for experimentation and measurements.  
 
In the mathematical modeling work, we model the impact on a surrogate vehicle of actions by malware as well as ``bonware'' -- the ensemble of cyber-physical features that defend the vehicle's computer system and allow it to recover from attack.  We develop a differential equation that models the effects of these competing forces on the system.  We extend this to a stochastic differential equation model that captures the effects of uncertainty and randomness in the activity times of malware and bonware.  
 
In parallel with the mathematical modeling, we have developed an inexpensive experimental environment to test the effects of malware and bonware on the Controller Area Network (CAN) bus of a generic military vehicle.  Our testbed includes (a)~a PASTA platform: a vehicle security testbed developed by Toyota which features a set of connected electrical control units or ECUs), (b)~Unity: a popular game development platform, and (c)~the Active Defense Framework (ADF): a government-developed framework used to quickly produce and test network-based cyber-defense techniques.  When running this environment, we capture data that can be used to characterize cyber resilience metrics of the modeled vehicles.
 
Although we have made significant progress in both of these areas, we have also identified some areas that may benefit from collaboration with research partners.  Thus far, we have modeled the impact on a single vehicle.  By considering a network of vehicles, many interesting problems can be formulated.  If one vehicle is under attack, what are the impacts on other vehicles that exchange information with this vehicle
or that are in physical proximity?
What are the probabilities that neighboring vehicles in the network are also under attack?  How can our understanding of vehicle networks and robotics inform each other?  What other techniques can we draw from to formulate models and compute metrics (for example game theory, neural networks, dueling network architectures, etc.)?

%% file: includes/20-prior-work-revisited.tex
\section{Prior work}\label{sec:prior-work}

In \cite{kott2022}, we reviewed literature related to qualitative and quantitative assessments of a cyber system.  We refer the reader to that paper for a more detailed summary.   Here, we briefly review the quantitative work to date, and motivate the need for stochastic modeling of cyber resilience.

\begin{figure}[ht]
  \centering
 \aaa{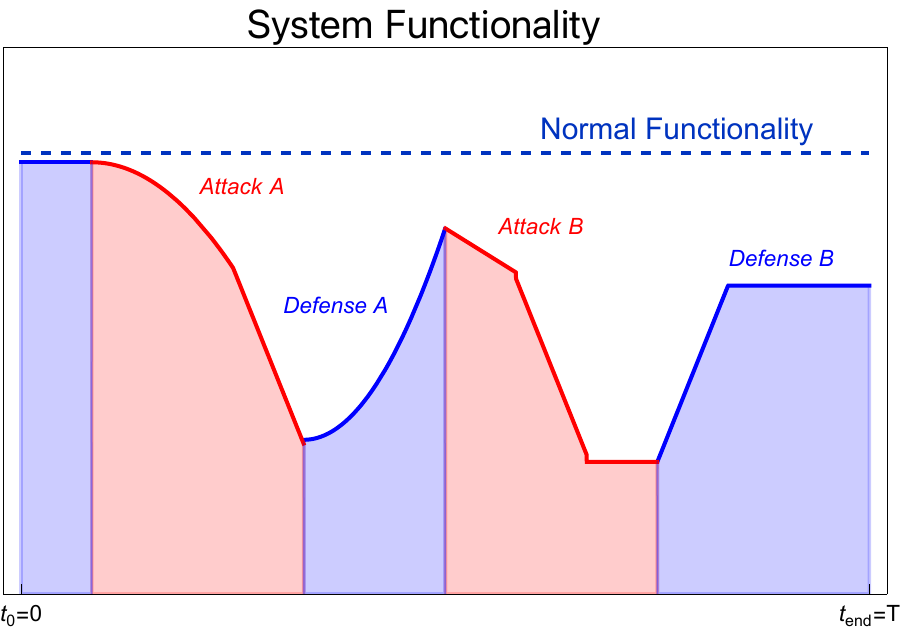}
 \caption{Resilience can be measured by subjecting a system to cyber attacks computing the ratio of the area under the compromised system functionality curve to that of the normal functionality curve.}
 \label{fig:auc}
\end{figure}

Most approaches to quantitative measurements of cyber resilience tend to involve the area under the curve (AUC) method \cite{hosseini2016review,kott2021to}.  An experimental system engages in a collection of missions where it subject to cyber attack.  Data containing the functionality of the system is collected, and a metric based on the area under the functionality curve is computed.  Figure~\ref{fig:auc} illustrates the concept.  One or more attacks compromise a system, causing the functionality to diminish.  The ratio of the area under the curve and the area under the baseline curve (recorded during a mission where there is no attack) is computed.  

This class of measures seems reasonable, however, AUC-based resilience measures are rather simple, and 
reveal little
about the underlying processes. In \cite{kott2022} we began to explore ways to quantify the resilience \impact{} of the bonware and quantify the \impact{} of malware on a system.  We also looked to understand how these values of \impact{fulness} vary over time during an incident.

%% file: includes/40-continuous-model-new.tex
\section{Continuous model}\label{sec:continuous}
\input{includes/40a-functionality}
 We assume that functionality is differentiable at least once and both
malware impact and bonware impact are continuous functions of time:
$\functionality \in C^1$ and $\malware, \bonware \in C^0.$
The \impact{} on functionality is the sum
of the \impact{s} of malware and bonware, and 
\begin{equation}
  \frac{d\functionality}{dt} + \Qware(t) \functionality(t) = \Fnominal \bonware (t),
\label{eq:00}
\end{equation}
where $\Qware(t)=\malware(t)+\bonware(t).$
In \cite{kott2022}, we reasoned that $\bonware(t), \malware(t) \ge 0$,
$\Fnominal>0$, and $\Fnominal \ge \functionality(t)  \ge 0$
and found the general solution to this first-order linear differential equation with
initial condition: 
\small
\begin{equation}\nonumber
  F(t)  = \expminus \left( F(0) + \bintegral \right).
\end{equation}
\normalsize
In \cite{kott2022} we also exhibited solutions for a number of elementary
examples.  Figure~\ref{fig:piecewise:pair} contains plots of piecewise constant and
piecewise linear models.  The models' differential equations and their solutions are summarized in Table~\ref{tab:models}.
\input{includes/table_one}

\input{includes/two_figures}

%% file: includes/40a-functionality.tex


We are interested in the functionality of our system, $\functionality(t)$, which we defined in \cite{kott2022} to be the time derivative of mission accomplishment.  We also proposed a baseline functionality (normal functionality), which in general could be time varying, but in our analysis, we take normal functionality, $\Fnominal(t) = \Fnominal$, to be a constant.  We also assume that the system, prior to any attack or other deviation from normal operations, is operating normally:  $F(t_0)=\Fnominal.$

%% file: includes/table_one.tex

\newcommand{\interval}{(t_j \le t < t_{j+1}), \quad (j=0, \cdots, N-1)}

\newcommand{\tempa}{\Omega(t)}
\newcommand{\tempaa}{\Omega(t-t_{j-1})}
\newcommand{\tempb}{\Lambda}

\newcommand{\tcj}{\Omega_j(t)}
\newcommand{\td}{\Lambda}

\newcommand{\tablelines}[3]{\begin{equation*} \text{#1} \end{equation*} & \begin{equation*}#2\end{equation*} & \begin{equation*}#3\end{equation*}\\}

\begin{table*}[htp]
\caption{Mathematical Models Developed in \cite{kott2022}}
\label{tab:models}
\begin{tabular}{ |p{2.25cm}|p{5.75cm}|p{8.75cm}|  }
\hline
\tablelines{}{\text{Differential Equation}}{\text{Solution}}
\hline
\tablelines{Continuous Model}{\frac{d\functionality}{dt} + \Qware(t) \functionality(t) = \Fnominal \bonware (t)}{  F(t)  = \expminus \left( F(0) + \bintegral \right)}
\hline
\tablelines{Constant}{\frac{d\functionality}{dt} + \Qware \functionality(t) =  \Fnominal \bonware}{\functionality(t) = \left[\functionality(0) - \frac{\Fnominal \bonware}{ \Qware } \right] e^{-\Qware t} + \frac{\Fnominal \bonware}{\Qware}}
\hline
  \tablelines{Piecewise Constant}
{  \begin{aligned}
& \frac{d\functionality}{dt} =
  \sum_{j=0}^{N-1}(\Fnominal-\functionality(t)) \bonware_j -
  \functionality(t) \malware_j,\\
&\interval
\end{aligned}}
{
  \functionality(t) = 
  \left[\functionality(t_{j-1}) - \frac{\Fnominal \bonware_j}{
      \Qware_j } \right]  e^{-\Qware_j (t-t_{j-1})} + \frac{\Fnominal
    \bonware_j}{\Qware_j} 
        }
\hline
\tablelines{Linear}{\frac{d\functionality}{dt} + (\lambda - \omega t)
                   \functionality(t) = \Fnominal (\alpha - \beta t)}{
  \begin{aligned}
    \frac{\functionality(t)}{\Fnominal} &= \frac{1}{\tempa} \left\{
      \frac{\functionality(0)}{\Fnominal}
      -\frac{\beta}{\omega}\left(1-{\tempa}\right) +(\alpha \omega
      -\beta \lambda )
    \right.\\
    \times&\left.  \frac{\sqrt{\frac{\pi }{2}} e^{\tempb^2} }{\omega
        ^{3/2}}\left[\erf\left(\tempb\right)+\erf\left(\frac{\omega
            t}{\sqrt{2 \omega }}-\tempb\right)\right] \right\}\\
    &\text{where } \Omega(t)=e^{\lambda t -\frac{1}{2}\omega t^2 }
    \text{ and } \td =\sfrac{\lambda }{\sqrt{2 \omega }}
            \end{aligned}}
\hline
      \tablelines{Piecewise Linear}{\begin{aligned}
          &\frac{d\functionality}{dt} = \sum_{j=0}^{N-1}
                   \left[ (\lambda_j - \omega_j t) \functionality(t) -
                   \Fnominal (\alpha_j - \beta_j t)
            \right],\\ &\interval \end{aligned} \label{eq:piecewise:linear} }
            {
            \begin{aligned}
              \frac{\functionality(t)}{\Fnominal} &= \frac{1}{\tempaa} \left\{
                \frac{\functionality(t_{j-1})}{\Fnominal}
                -\frac{\beta}{\omega}\left(1-{\tempaa}\right) +(\alpha \omega
                -\beta \lambda )
              \right.\\
              \times&\left.  \frac{\sqrt{\frac{\pi }{2}} e^{\tempb^2} }{\omega
                  ^{3/2}}\left[\erf\left(\tempb\right)+\erf\left(\frac{\omega
                      (t-t_{j-1})}{\sqrt{2 \omega }}-\tempb\right)\right]
              \right\},\\
              &\text{where } \tcj=e^{\lambda_j (t-t_j)-\frac{1}{2}\omega_j (t-t_j)^2 }
              \text{ and } \td_j=\sfrac{\lambda_j }{\sqrt{2 \omega_j }}
            \end{aligned}
                }
\hline
\end{tabular}
\end{table*}

%% file: includes/two_figures.tex
  
\begin{figure*}[ht] 
  \centering
  \aaa{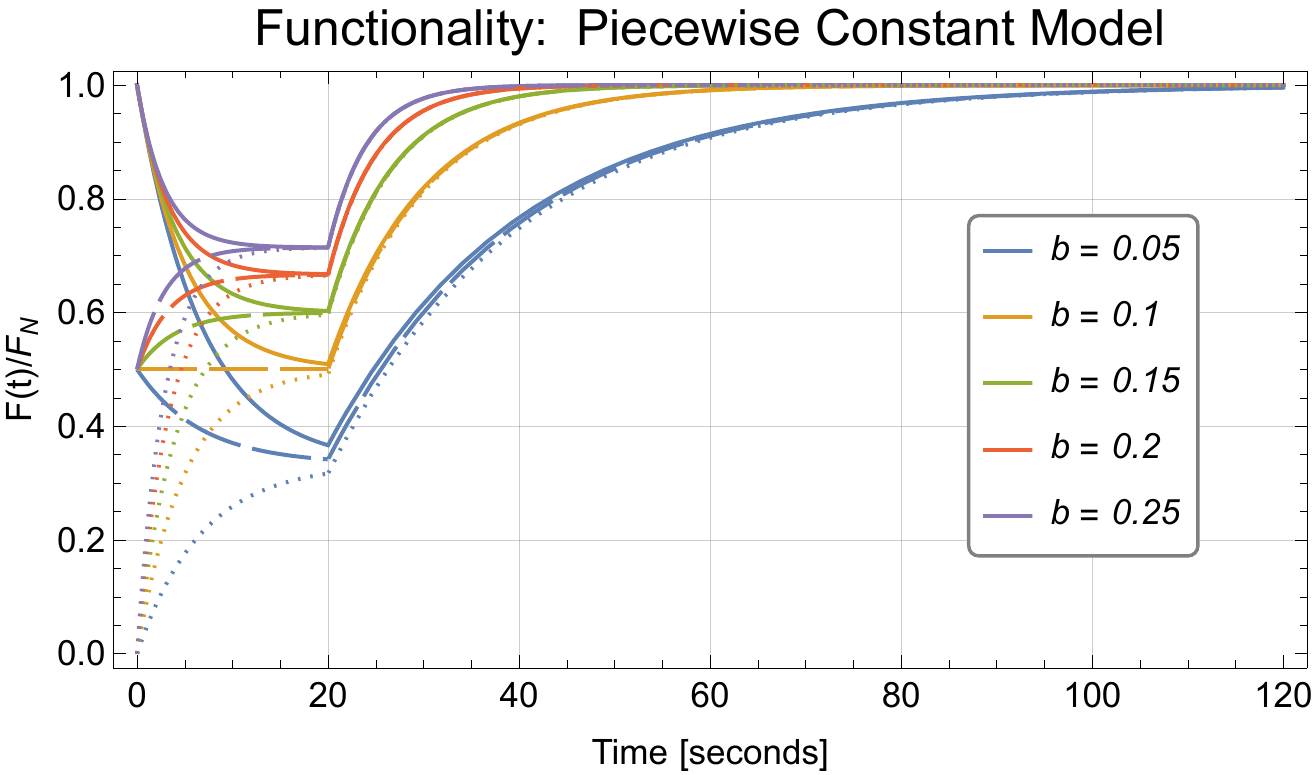}
  \aaa{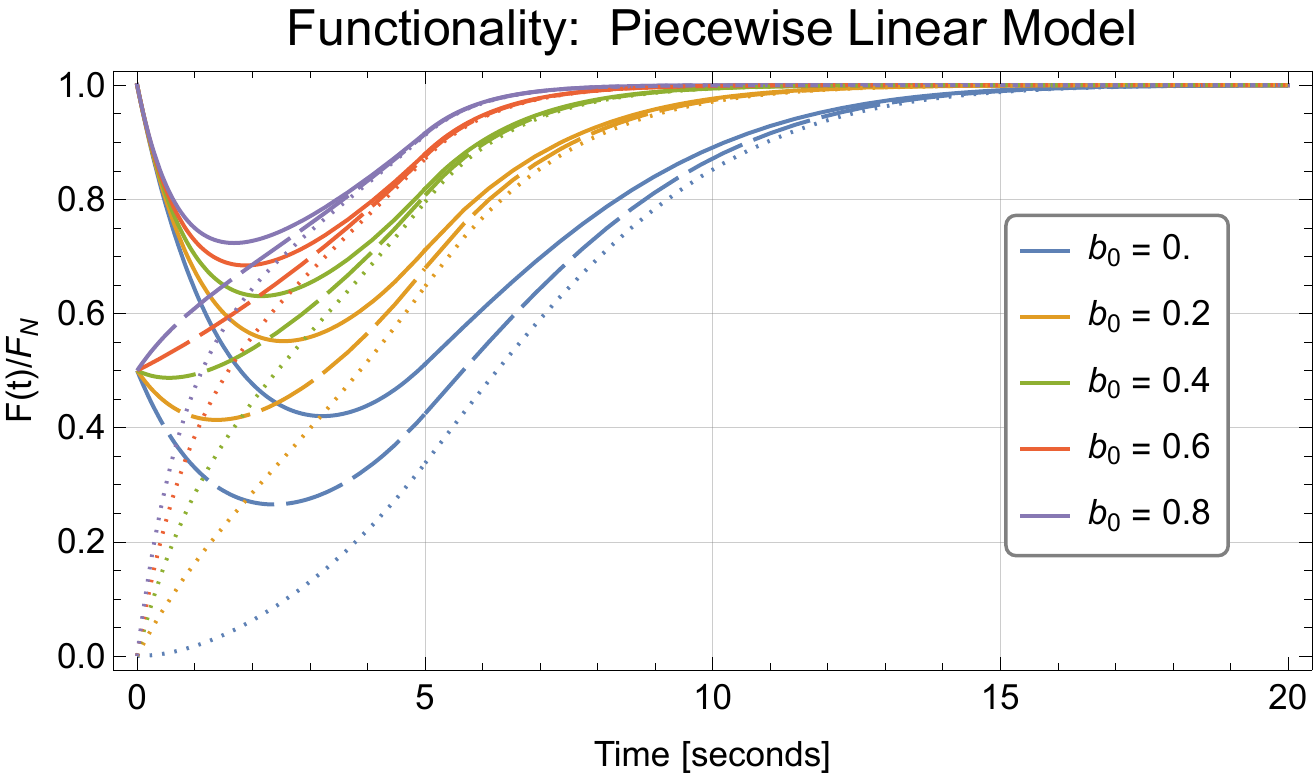}
  \caption{Left: Normalized functionality,
    $\functionality(t)/\Fnominal$, is shown for piecewise constant
    models (Row 3 in Table 1).  The malware impact is
    $\malware(t)=[u(t)-u(t-20)]$ and the bonware impact is
    $\bonware=b u(t-20).$ Initially, malware attacks at mission time
    $t=0.$ At time $t=20$ seconds, bonware becomes aware of the
    attack, counters malware, and brings the system back towards
    normal operations.  Right: Normalized functionality,
    $\functionality(t)/\Fnominal$, is shown for piecewise linear
    models (Row 5 in Table 1). The malware impact is
    $\malware(t)=\max (0,0.5-0.1t)$ and the bonware impact is
    $\bonware=b_0+0.04t.$ Both malware and bonware \impact{}s are
    initially linear functions of time.  When malware \impact{}
    reaches $\malware=0$, then malware \impact{} is zero but bonware
    \impact{} continues to increase.  The function $u(t)$ is the unit
    step function: $u(t)=0$ when $t<0$ and $u(t)=1$ when $t\ge 0.$ The
    figure on the right is from \cite{kott2022}, used by permission.}
   \label{fig:piecewise:pair}
\end{figure*}

%% file: includes/figure_3.tex
\begin{figure*}[t]
  \centering
  \eee{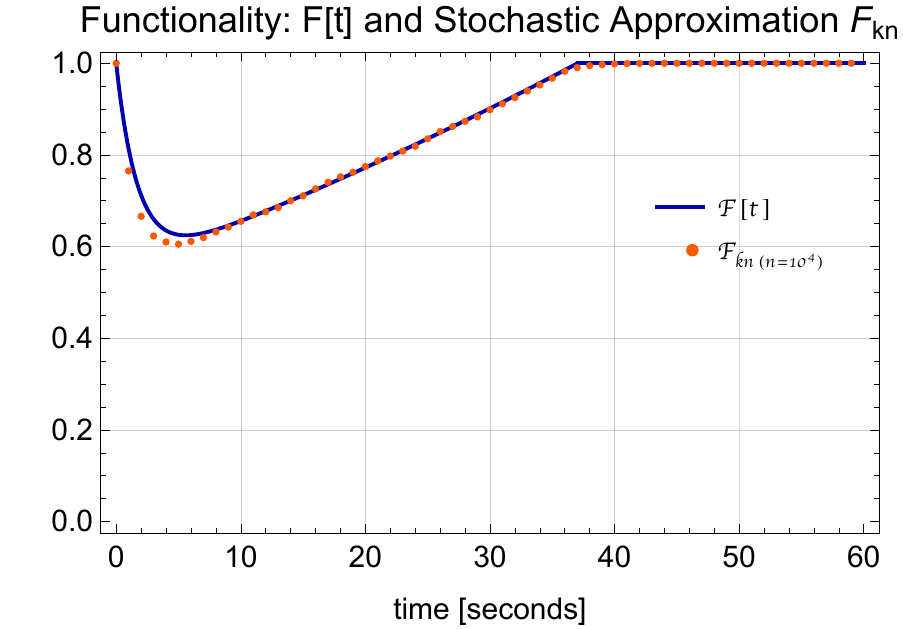} 
 \hspace{1mm}
  \aaa{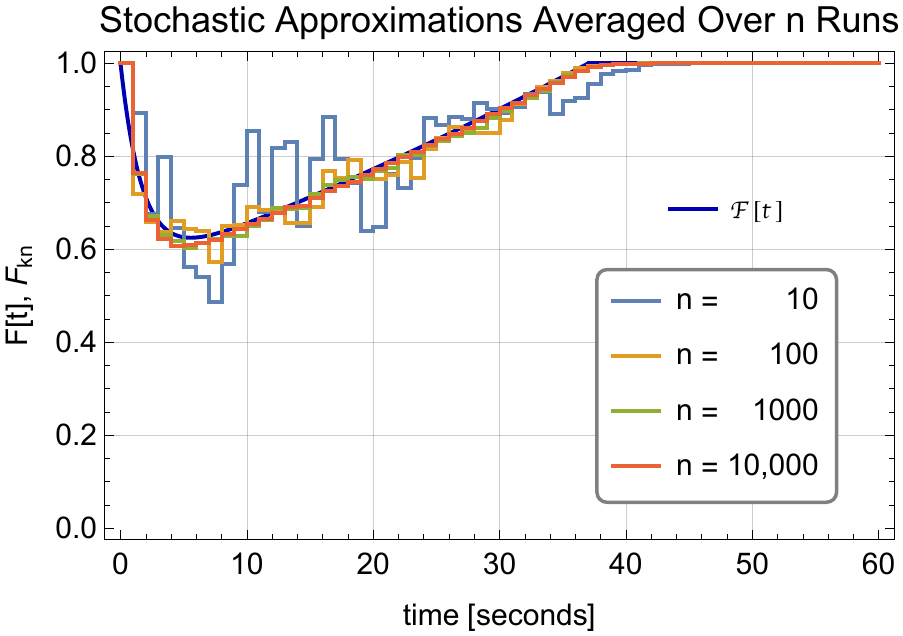}  
   \caption{Left:  The functionality for a piecewise linear model is
      shown in blue.
      $\malware(t)=\max(0,0.25-0.00675 t)$ and $\bonware(t)~=~\min(0.5,
      0.3~+~0.0045 t).$  The average of 10,000 runs for the
      corresponding stochastic model is shown in orange.  Right:  The
      average of $n\in\{10,100,1000,10,000\}$ runs.  As the number of
      runs averaged increases, the aggregate stochastic model
      approaches the solution of the continuous differential equation.}
    \label{fig:3}
\end{figure*}

%% file: includes/50-stochastic-differential-equation-model-new.tex
\section{Stochastic differential equation model}

In this section, we develop the \textit{stochastic} differential equation (SDE) model associated with the piecewise linear model that we introduced in \cite{kott2022} (see Table~\ref{tab:models}).  The extension is motivated by the discontinuous nature of the notional data in Figure~\ref{fig:auc}.  Whereas the differential equation model assumed a smooth functionality curve, our stochastic version allows for a more punctuated attack-and-restoration pattern.  
In \cite{kott2022} we obtained 
\begin{equation}\label{eq:sde:generic}\nonumber
  \frac{d \functionality}{d t} 
  =  \left(\Fnominal-\functionality(t)\right) \bonwareActivity \bonwareEffectiveness - \functionality(t) \malwareActivity \malwareEffectiveness.   
\end{equation}
which we approximate by the stochastic difference equation
\begin{equation}\label{eq:stochastic:solution}
  F_{k}=F_{k-1}+A^b(k) E^b(k)(F_N-F_{k-1}) - A^m(k) E^m(k) F_{k-1}
\end{equation}
with parameters
\begin{eqnarray*}
  \malwareActivity      &\sim& \dbern{\malwareRate(t)}\label{eq:sde:mwa}, \\
  \bonwareActivity      &\sim& \dbern{\bonwareRate(t)}\label{eq:sde:bwa}, \\
  \malwareEffectiveness &\sim& \dunif{0}{\malwareSize(t)}\label{eq:sde:mwe},\\
  \bonwareEffectiveness &\sim& \dunif{0}{\bonwareSize(t)}\label{eq:sde:bwe},
\end{eqnarray*}
where $\dbern{\theta}$ indicates the Bernoulli distribution with rate $\theta$ and $\dunif{0}{\gamma}$ indicates a uniform distribution with lower bound $0$ and upper bound $\gamma$.


Hence, $\malwareRate(t) \in {[0,1]}$ is the probability that malware is successful at time $t$, $\bonwareRate(t) \in {[0,1]}$ is the probability that bonware is successful at time $t$, $\malwareSize(t) \in {(0,1]}$ is the maximum fraction of damage inflicted by malware, and $\bonwareSize(t) \in {(0,1]}$ is the maximum fraction of damage undone by bonware.

Like the ordinary differential equation (ODE) model, the SDE model allows for a number of interesting variants.  In the remainder of this section, we introduce its extension to piecewise linear models.

%% file: includes/52a-piecewise-linear-sde.tex
\subsection{Piecewise linear parameters}


%
 In the plot on the left of Figure 2, malware is proportional to the
 difference of two step functions (malware is active starting from
 time $t=0$ seconds and is turned off starting at time $t=20$ seconds.
 Various levels of bonware are depicted.  Each is expressed as a
 weighted step function, which turns on at time $t=20$ seconds.  In
 the plot on the right, malware impact is originally at $\malware(0)=0.5$
 and decreases over time, but we enforce $\malware \ge 0.$
 Various linear functions of bonware are shown.  In both figures,
 curves with initial conditions of $\{0,0.5,1.0\}$ are illustrated.

In the example depicted in Figure 3, we set:
\begin{equation*}
  \begin{aligned}
  \malwareRate(t) & = 2\malware(t) = \max(0,0.5 - 0.135 t), \\
  \bonwareRate(t) & = 2\bonware(t) = \min(1,0.6+0.009 t).
\end{aligned}
\end{equation*}

%% file: includes/60-relationship-between-ode-and-sde-new.tex
\subsection{Relationship between continuous and SDE model}\label{sec:proposition}

With the parameters of the stochastic model selected appropriately, we
showed in \cite{kott2022} that as the number of stochastic
realizations increases, the expectation of the solution to the
stochastic differential equation model approaches that of the ODE
model.  In \cite{kott2022} we proved the following theorem:

\newtheorem*{theorem}{Theorem}

\begin{theorem}
Let $y^m_k \sim \dbern{2\malware}$, $y^b_k \sim
\dbern{2\bonware}$, $z^m_{k} \sim \dunif{0}{\functionality_{k}}$, $z^b_{k} \sim \dunif{0}{\Fnominal-\functionality_{k}}$, and 
\begin{equation*}  \label{eq:stochastic:diff}
  \functionality_{k+1}  =  \functionality_{k} 
     - y^m_{k} z^m_{k}
     + y^b_{k} z^b_{k}, \quad (k=1,\hdots,K). 
\end{equation*}
Let ${\mathcal{F}_k}_n=\frac{{\functionality_k}_j}{n}$, $(j=1,\hdots,n)$, then\\
\blue{$\mathcal{F}_k~=~\mathbb{E} (\functionality_k)~=~\lim_{n\to\infty}
 {\mathcal{F}_k}_n$ and $\mathcal{F}_k\approx \functionality(k),$ for
 large $k$, 
 where $\functionality(t)$ is the solution to the initial value problem
given by Equation~\ref{eq:00} with $F(0)=\mathcal{F}_0.$}
\end{theorem}


%% file: includes/70a-generating-stochastic-realizations.tex
\section{Generating stochastic realizations}
In \cite{kott2022} we developed a method to extract the parameters of
the stochastic model associated with a continuous model.  In the left
pane of Figure~\ref{fig:3},
we plot a continuous model along with the average over 10,000
runs of its associated stochastic model.  After the first two or three
initial points, the agreement is excellent.  In the right pane of
Figure~\ref{fig:3}, we show averages of $n$ instantiations of the stochastic
model, for $n\in \{10,100,1000, 10,000\}.$  As $n$ increases, we see
better agreement with the solution to the continuous differential
equation model.  In Figure~\ref{fig:4}, we plot the absolute error between
the two models.  After the first five seconds of data, the absolute
error between the continuous model solution and the average of the
stochastic ensemble is less than 0.01.

%% file: includes/figure_4.tex
\begin{figure}[t] 
  \centering
 \aaa{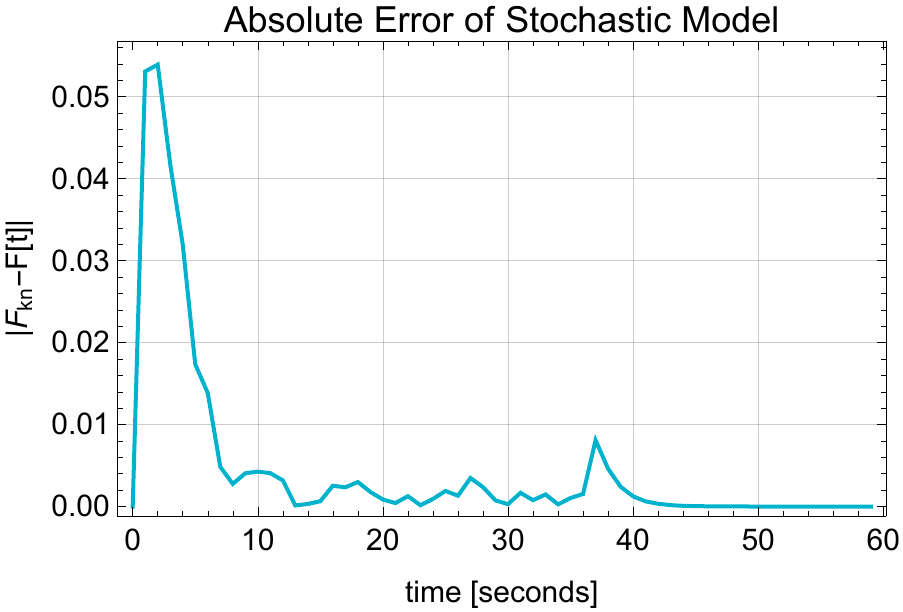}
    \caption{The absolute error of the stochastic model calculated
      as the absolute value of the difference between the solution to the
      differential equation and the average of 10,000 runs.}
    \label{fig:4}
\end{figure}

%% file: includes/75-parameter-estimation.tex
\section{Parameter estimation}
Given the deterministic solutions to the stochastic differential equations presented above (see Eq.~\ref{eq:stochastic:solution}), we can now proceed to the estimation of model parameters from data.

The parameter estimation has two components: (a) a loss function that describes the distance between the observed data $Y$ and the data implied by the model with parameter vector $\Theta$, and (b) an optimization routine to find those values for $\Theta$ that minimize the loss.

We considered two loss functions, depending on whether the performance data are constrained to a fixed domain or not.  First, we considered the case where our performance data is expressed as a fraction of optimal performance and constrained to the open $(0, 1)$ interval.  For such cases, a convenient option is the \emph{beta 
loss} function \cite{hastie}:
\begin{eqnarray*}\nonumber
&L_\beta\left(\Theta\right) =&  -\sum_k F_k\left(\Theta\right) \log\left(Y_k\right) \\
&& \quad-\sum_k \left(1-F_k\left(\Theta\right)\right)\log\left(1-Y_k\right),\qquad\qquad
\end{eqnarray*}
where $F_k\left(\Theta\right)$ is the model-predicted performance at the $k^\text{th}$ observation, given parameter vector $\Theta$; and $Y_k$ is the $k^\text{th}$ observed performance value.  $F_k\left(\Theta\right)$ can be calculated as the deterministic solution to the SDE as in Equation~\ref{eq:stochastic:solution}, if that solution is available.  Alternatively, if such a solution is unavailable or cumbersome to compute, $F_k\left(\Theta\right)$ could be numerically approximated using an iterative method such as the forward Euler method or other Runge-Kutta methods \cite{press2007numerical}.

However, in our applications, the data are not usually restricted to this limited domain -- nominal performance levels are often unknown or variable, and when they are known, performance is often at 100\%.  Therefore, in practice, we will use a \emph{squared error loss} function \cite{hastie}:
$$L_s\left(\Theta\right) = \sum_k \left(Y_k-F_k\left(\Theta\right)\right)^2.$$

The parameter vector $\Theta$ itself usually contains at least some elements that are restricted to a limited domain (e.g., the bonware and malware effectiveness parameters are strictly positive).
This leaves us with a complex constrained optimization problem.  Fortunately, there exist several well-known algorithms that can quickly find optima of loss functions under constraint, especially if the dimensionality of $\Theta$ is low, as it is here.  We opted for a Nelder-Mead simplex optimization algorithm \cite{nelder}, which is fast, robust, and easy to implement.  The Nelder-Mead simplex procedure is implemented in MATLAB's \texttt{fminsearch}, R's \texttt{optim}, and Python's \texttt{scipy.minimize}.
The optimization procedure will yield the estimated parameter vector $\hat{\Theta} = \text{arg min}_\Theta L_s\left(\Theta\right)$.

%% file: includes/76-next-steps.tex
\section{Next steps}
The next steps in the development of our technology for quantifying cyber resilience will be to test the efficiency and precision of these estimation models through numerical experiments (i.e., simulation studies).  While the methods we use are well established, their application to the piecewise linear SDE models is not.

\begin{figure}[h]
    \centering
    \aaa{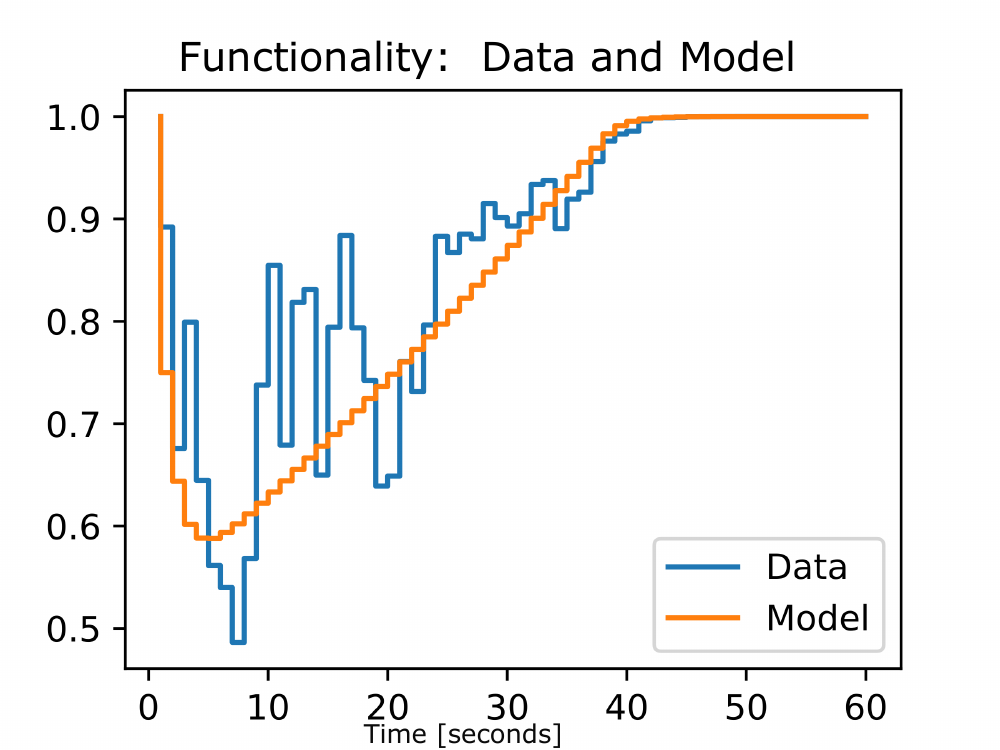}
    \caption{
    Model fit to the average of 10 runs.
    The model captures the general trend of the data well despite the fact that estimated parameters are quite different from the generating parameters.
    }
    \label{fig:piecewise_linear_fit}
  \end{figure}

One specific issue to examine is that of \textit{mimicry} -- a phenomenon where multiple distinct combinations of parameters yield predictions that are impossible to distinguish with the available amounts of data.  Figure~\ref{fig:piecewise_linear_fit} illustrates this issue.  Here, we generated 10 runs of 60 observations using the same parameters as those used to generate Figure~\ref{fig:3}: $\malware(t)=\max\{0, m_0 - m_1 t\}$ and $\bonware(t) = \min\{0.5, b_0 + b_1 t\}$ with $m_0 = 0.25$, $m_1 = 0.00675$, $b_0 = 0.3$, and $b_1 = 0.0045$.  However, the estimated parameters were $\hat{m}_0 \approx 0.3146$, $\hat{m}_1 \approx 0.5227$, $\hat{b}_0 \approx 0.5447$, and $\hat{b}_1 \approx 0.5568$.  Despite these clear differences, the model captures the fast-downward-then-slow-upward pattern in the data well.  This indicates that the parameters of this model may be difficult to identify given the other qualities of the data (such as the sample size and magnitude of the random variability) or certain specific features of the generating parameter sets (e.g., the relative timing of the effective onset of malware and bonware).

%% file: includes/77-linear-filtering.tex
\section{Linear Filtering}
Obtaining an initial approximation of the parameters of the stochastic difference equation (2), and positing an estimate of their uncertains, we can apply linear filtering theory, and frame our estimation problem as a Kalman-Bucy filter \cite{jazwinski}.  Our piecewise linear model includes abrupt changes in the trajectories of both malware and bonware.  These changes can be accounted for by allowing random maneuvers in the ``target dynamics'' \cite{brookner} or by employing interacting multiple models (IMM)  \cite{blackman} \cite{barshalom} where multiple filters track the parameters and a mechanism is established to choose which filter is most appropriate at each time increment \cite{willsky}.  For each of a finite set of parameters governing both malware and bonware impact, a filter may be established, and transitions can be tracked by evaluating the relative performance of these filters.  A related approach \cite{willskyandjones} is to view our system as a linear stochastic system with unknown jumps (changes in the linear models governing the malware and bonware impacts).  Since the changes occur infrequently, a monitoring system is set up to monitor filter residuals.  When these become large, an adjustment is made to the filter.

%% file: includes/80-discussion-and-conclusion-new.tex
\section{Discussion and conclusion}

In \cite{kott2022}, we have presented a broadly applicable framework for the analysis of the cyber resilience of military artifacts.  Our framework relies on the construction of a custom differential equation time series model that shows good qualitative correspondence to the functionality of vehicles performing missions.  


Both types of models can be extended to a large variety of custom circumstances, including the case where model parameters change gradually, abruptly, or predictably as a result of experimental manipulation.  In this paper, we have extended the stochastic model to include piecewise linear malware and bonware activities.  The piecewise linear model is mathematically tractable and can be used to generate data that qualitatively correspond to performance data in our lab tests.  However, further analysis is needed to establish the conditions under which parameters of the model can be reliably estimated.